\documentclass[eqsecnum,superscriptaddress,nofootinbib,11pt]{revtex4}
\usepackage{amsmath}
   \usepackage{bm}
\usepackage{amsthm,amscd}
\usepackage{mathrsfs}
\usepackage{verbatim}
\usepackage{amsfonts,amsmath,latexsym,amssymb,txfonts,bm}
\usepackage[american]{babel}
\usepackage{dsfont}
\usepackage[dvips]{graphicx}
\usepackage{latexsym}
\usepackage{epsfig}
\usepackage{braket}
\usepackage{color}
\newcommand{\diff}{\mathrm{d}}

\def\beq{\begin{eqnarray}}
\def\eeq{\end{eqnarray}}

\begin{document}
\title{The Casimir effect for pistons with transmittal boundary conditions}

\author{Guglielmo Fucci\footnote{Electronic address: fuccig@ecu.edu}}
\affiliation{Department of Mathematics, East Carolina University, Greenville, NC 27858 USA}


\date{\today}
\vspace{2cm}
\begin{abstract}

This work focuses on the analysis of the Casimir effect for pistons subject to transmittal boundary conditions.
In particular we consider, as piston configuration, a direct product manifold of the type $I\times N$ where $I$ is
a closed interval of the real line and $N$ is a smooth compact Riemannian manifold. By utilizing the spectral zeta function
regularization technique, we compute the Casimir energy of the system and the Casimir force acting on the piston.
Explicit results for the force are provided when the manifold $N$ is a $d$-dimensional ball.

\end{abstract}
\maketitle

\section{Introduction}

The Casimir effect, which was first predicted theoretically in 1948 as an attraction between perfectly conducting neutral plates
\cite{casimir}, is indisputably one of the most studied subjects in the ambit of quantum field theory
and its interaction with external conditions. The analysis of the Casimir effect has evolved from the simple
configuration of two parallel plates considered in the pioneering work of Casimir, to much more complex geometries and boundary conditions
(the reader may refer, for instance, to \cite{bordag01,bordag09,milton01,plunien86} and references therein).
As a result of the very nature of the Casimir effect, which is a set of phenomena originating from modifications to the vacuum energy of a quantum field
due to its interaction with external conditions, calculation of the Casimir energy of a quantum field often lead to a divergent quantity.
In order to extract meaningful results for the Casimir energy, a suitable regularization procedure is required. In this work we employ the spectral zeta
function regularization technique \cite{blau88,bytsenko03,elizalde94,elizalde,kirsten01,kirsten10} which represents one of
the most widely used regularization methods.

Piston configurations, which were first introduced by Cavalcanti in 2004 \cite{caval04}, have lately been the main focus of
a large body of research. The reason for this widespread interest is encoded in the particular geometry that piston configurations have.
In the most general case, a piston configuration consists of two $D$-dimensional manifolds, often referred to as chambers, with a
common boundary being a codimension one manifold representing the piston.
One of the main portions of the analysis of piston configurations consists in the computation of the Casimir force acting on the piston.
This force is the result of differences in vacuum energy in the two chambers. What makes pistons such an attractive geometric configuration
for studying the Casimir effect is the fact that while their Casimir energy might be divergent, the force acting on the piston itself is,
in most cases, well defined. In fact, there exist piston configurations with non-vanishing curvature in which the Casimir force acquires divergent terms
that are proportional to particular geometric invariants of the piston \cite{fucci11,fucci11b,fucci12}.
The Casimir effect has been studied throughout the literature for a plethora of piston configurations, a small sample of this
work can be found, for instance, in \cite{barton06,edery06,hertz07,li97,marachevsky07,kirsten09}. Most of the investigations
regarding the Casimir effect for pistons are developed by assuming that the quantum field propagating in the piston
satisfies ordinary boundary conditions, such as Dirichlet, Neumann, Robin, or mixed. Although of enormous theoretical importance, the aforementioned
boundary conditions only model idealized situations. In order to overcome this limitation, piston configurations have been analyzed
for quantum fields endowed with more general self-adjoint boundary conditions (see e.g. \cite{asorey13,fucci15}) in an attempt to describe physical, as opposed to idealized,
situations. Another approach which has been utilized to imitate physical boundary conditions is based on the replacement of the
piston itself and the associated boundary conditions with a smooth potential with compact support \cite{bea13}.
Many of the methods employed to analyze the Casimir effect for piston configurations have a specific characteristic in common.
In describing the propagation of quantum fields in the piston configuration, one is led to solve an eigenvalue equation
endowed with particular boundary conditions in each chamber \emph{separately}. In this way the behavior of the quantum field in one chamber is completely
independent of its behavior in the other chamber. In other words, ordinary boundary conditions imposed on the piston do not allow
any interaction between the two chambers of the piston. These approaches are clearly unsuitable if one wishes to consider
configurations possessing, for example, a semi-transparent piston where the quantum field propagating in one chamber influences
the propagation of the field in the other chamber through the piston. In order to allow interaction between the two chambers
one could impose transmittal boundary conditions on the piston \cite{gilkey01,gilkey03,gilkey04,kirsten01,kirsten02}.

In this work we focus on the analysis of the Casimir effect for a scalar field propagating in a piston configuration of the type $I\times N$ where
$I$ is a closed interval of the real line and $N$ is a smooth compact Riemannian manifold.
On the piston itself we impose transmittal boundary conditions and utilize the spectral zeta function regularization technique to obtain the
Casimir energy of the system and the associated force on the piston. As we will observe in the next section, transmittal boundary conditions
are dependent on a parameter. One of the main goals of this work is to show how the Casimir force on the piston depends on the parameter describing the transmittal boundary conditions.

The outline of the paper is as follows. In the next section we describe in details the spectral zeta function associated with the piston configuration 
and the boundary conditions that are considered in this work. Section \ref{sec3} contains an outline of the analytic continuation of the spectral zeta function 
which is necessary for the computation of the Casimir energy and force. In Section \ref{sec4} we analyze the case in which the Laplacian on the piston possesses 
a zero mode and compute its contribution to the spectral zeta function. In Section \ref{sec5} we evaluate the Casimir energy and force for 
the piston configuration and in Section \ref{sec6} we utilize the results to compute numerically the force when the piston is a $d$-dimensional ball.
The last section points to the main results of this paper and outlines possible further studies related to Casimir pistons with transmittal boundary conditions.

\section{The spectral zeta function and Casimir energy}

We consider a generalized piston configuration, analyzed also in \cite{fucci12,fucci15}, based on a product
manifold of the type $M=I\times N$ where $I=[0,L]\subset\mathbb{R}$ is a closed interval
of the real line and $N$ is a smooth compact Riemannian manifold with or without a boundary $\partial N$. We
assume, furthermore, that the dimension of the base manifold $N$ is $d$ which implies that $M$ is, hence, a manifold of dimension $D=d+1$.
The piston configuration can be constructed from the product manifold $M$ by placing at any point $a\in(0,L)$ the manifold $N_{a}$ which represents
the cross-section of $I\times N$ at $a$. In doing so, the manifold $M$ is divided in two chambers, denoted by $M_{I}$ and $M_{II}$,
separated by the common boundary $N_{a}$ which describes the piston itself. The two chambers, by construction, are smooth compact Riemannian manifolds with a boundary where
$\partial M_{I}=N_{0}\cup N_{a}\cup \left([0,a]\times N\right)$ and $\partial M_{II}=N_{a}\cup N_{L}\cup \left((a,L]\times N\right)$.

We consider a massless scalar field $\phi$ propagating on $M$. By using a set of coordinates $(x,\bf{X})$ where $x\in I$ and $\bf{X}$ represent the coordinates on $N$, the dynamics of the scalar field is described by the eigenvalue equation
\begin{equation}\label{0}
-\left(\frac{\diff^{2}}{\diff x^2}+\Delta_{N}\right)\phi=\alpha^2\phi\;,
\end{equation}
where $\Delta_{N}$ denotes the Laplacian on the manifold $N$ and $\phi$ belongs to the space $\mathscr{L}^{2}(M)$ of square integrable functions on $M$.
The eigenvalues $\alpha$ are uniquely determined once appropriate boundary conditions
are imposed on the scalar field $\phi$. The general solution to the differential equation (\ref{0}) can be written as a product $\phi=f(x)\varphi\left(\bf{X}\right)$ where
$\varphi\left(\bf{X}\right)$ are the eigenfunctions of the Laplacian on $N$ with eigenvalues $\lambda$, that is
\begin{equation}\label{1}
-\Delta_{N}\varphi=\lambda^{2}\varphi\;,
\end{equation}
and the function $f(x,\lambda)$ satisfies the following second-order ordinary differential equation
\begin{equation}\label{2}
\left(-\frac{\diff^2}{\diff x^2}+\lambda^{2}-\alpha^{2}\right)f_{\lambda}(x,\alpha)=0\;.
\end{equation}
The eigenvalues $\alpha$ are, then, used to construct the spectral zeta function associated with our system as follows:
\begin{equation}\label{3}
\zeta(s)=\sum_{\alpha}\alpha^{-2s}\;,
\end{equation}
which, according to the general theory of spectral functions \cite{elizalde94, elizalde,kirsten01}, is well defined in the region of the complex plane
$\Re(s)>D/2$. In the ambit of the spectral zeta function regularization method, the function defined in (\ref{3}) can be utilized
to compute the Casimir energy of the system \cite{bordag01,bordag09,bytsenko03,elizalde94,elizalde,kirsten01}.
After performing a suitable analytic continuation,
$\zeta(s)$ in (\ref{3}) can be extended to a meromorphic function in the entire complex plane possessing only simple poles.
The analytically continued expression is then employed to express the Casimir energy as
\begin{equation}\label{4}
E_{\textrm{Cas}}(a)=\lim_{\epsilon\to 0}\frac{\mu^{-2s}}{2}\zeta\left(\epsilon-\frac{1}{2},a\right)\;,
\end{equation}
where $\mu$ is a parameter with the dimension of mass.
Since $\zeta(s)$ generally presents a pole at $s=-1/2$, the limit in (\ref{4}) leads to
\begin{equation}\label{5}
E_{\textrm{Cas}}(a)=\frac{1}{2}\textrm{FP}\,\zeta\left(-\frac{1}{2},a\right)+\frac{1}{2}\left(\frac{1}{\epsilon}+\ln\mu^{2}\right)\textrm{Res}\,\zeta\left(-\frac{1}{2},a\right)+O(\epsilon)\;.
\end{equation}
The force acting on the piston positioned at $x=a$ is given by
\begin{equation}\label{6}
F_{\textrm{Cas}}(a)=-\frac{\partial}{\partial a}E_{\textrm{Cas}}(a)\;.
\end{equation}
It is clear, from (\ref{5}) and (\ref{6}), that the Casimir force on the piston is free from divergences, and hence well defined, if
the residue of the spectral zeta function at $s=-1/2$ is independent of $a$.

As we have previously mentioned, the eigenvalues $\alpha$ in (\ref{0}) can be uniquely determined once appropriate boundary conditions are satisfied.
In this work, we impose ordinary boundary conditions, namely Dirichlet and Neumann, at the two end-points of the piston configuration $N_{0}$ and $N_{a}$.
The piston itself $N_{a}$ is, instead, endowed with transmittal boundary conditions which can be described as following \cite{gilkey01,gilkey03,kirsten02}:
Let $\mathcal{M}=\mathcal{M}_{I}\cup_{\Sigma} \mathcal{M}_{II}$ be a $d$-dimensional compact Riemannian manifold with $\Sigma$ being a codimension-one common boundary.
If $D_{I}$ and $D_{II}$ denote the Laplace operators on $\mathcal{M}_{I}$ and $\mathcal{M}_{II}$ acting on $\phi_{I}\in V|_{\mathcal{M}_{I}}$
and $\phi_{II}\in V|_{\mathcal{M}_{II}}$, respectively, then the transmittal boundary conditions are defined as $\mathcal{B}_{U}\phi=0$ with the transmittal operator being
\begin{equation}\label{7}
\mathcal{B}_{U}\phi=\left\{\phi_{I}|_{\Sigma}-\phi_{II}|_{\Sigma}\right\}\oplus\left\{\left(\nabla_{m_{I}}\phi_{I}\right)|_{\Sigma}+\left(\nabla_{m_{II}}\phi_{II}\right)|_{\Sigma}+U\phi_{I}|_{\Sigma}\right\}\;.
\end{equation}
In the above expression $U$ is an endomorphism of $V|_{\Sigma}$ and $\nabla_{m_{I}}$ and $\nabla_{m_{II}}$ are the exterior normal derivatives on $\mathcal{M}_{I}$ and $\mathcal{M}_{II}$, respectively, to the boundary $\Sigma$.

In order to construct the spectral zeta function of our system we need to know the eigenvalues $\alpha$. Although the eigenvalues cannot be explicitly
found in general, the boundary conditions will provide implicit equations to obtain them. The general solution to the ordinary differential equation (\ref{2}) in the first chamber
$M_{I}$ is a simple linear combination of trigonometric functions
\begin{equation}\label{8}
f_{I,\lambda}(x,\alpha)=a_{I}\sin\left(\sqrt{\alpha^{2}-\lambda^{2}}\,x\right)+b_{I}\cos\left(\sqrt{\alpha^{2}-\lambda^{2}}\,x\right)\;.
\end{equation}
In the second chamber, namely $M_{II}$, we find a similar general solution to (\ref{2}) which can be written, for later convenience, as
\begin{equation}\label{9}
f_{II,\lambda}(x,\alpha)=a_{II}\sin\left[\sqrt{\alpha^{2}-\lambda^{2}}(L-x)\right]+b_{II}\cos\left[\sqrt{\alpha^{2}-\lambda^{2}}(L-x)\right]\;.
\end{equation}
The general solutions (\ref{8}) and (\ref{9}) are then required to satisfy transmittal boundary conditions $\mathcal{B}_{U}\phi=0$ in (\ref{7}) at the piston $N_{a}$, namely
\begin{eqnarray}\label{10}
f_{I,\lambda}(a,\alpha)&=&f_{II,\lambda}(a,\alpha)\nonumber\\
f'_{I,\lambda}(a,\alpha)&=&f'_{II,\lambda}(a,\alpha)+U f_{I,\lambda}(a,\alpha)\;,
\end{eqnarray}
with the prime indicating differentiation with respect to the variable $x$,
together with a set of boundary conditions at the two endpoints $N_{0}$ and $N_{L}$ of the piston configuration.
In this work we consider three types of boundary conditions. The first ones, which we denote by the name \emph{Dirichlet-Dirichlet} are
\begin{equation}\label{11}
f_{I,\lambda}(0,\alpha)=f_{II,\lambda}(L,\alpha)=0\;.
\end{equation}
The subsequent conditions take the form
\begin{equation}\label{12}
f'_{I,\lambda}(0,\alpha)=f'_{II,\lambda}(L,\alpha)=0\;,
\end{equation}
which we designate as \emph{Neumann-Neumann} boundary conditions. The last set of boundary conditions, which we call \emph{mixed},
fall into two subsets, defined by the equations
\begin{equation}\label{13}
f_{I,\lambda}(0,\alpha)=f'_{II,\lambda}(L,\alpha)=0\;,\quad\textrm{and}\quad f'_{I,\lambda}(0,\alpha)=f_{II,\lambda}(L,\alpha)=0\;.
\end{equation}
In the fist case one imposes Dirichlet conditions at $N_{0}$ and Neumann ones at $N_{L}$ while in the second case the conditions are reversed.

By imposing Dirichlet-Dirichlet boundary conditions (\ref{11}) and the transmittal boundary conditions (\ref{10}) on the general solutions (\ref{8})
and (\ref{9}) we obtain $b_{I}=b_{II}=0$ and the linear system
\begin{eqnarray}\label{14}
a_{I}\sin\left(\sqrt{\alpha^{2}-\lambda^{2}}\,a\right)&=&a_{II}\sin\left[\sqrt{\alpha^{2}-\lambda^{2}}(L-a)\right]\nonumber\\
a_{I}\sqrt{\alpha^{2}-\lambda^{2}}\cos\left(\sqrt{\alpha^{2}-\lambda^{2}}\,a\right)&=&-a_{II}\sqrt{\alpha^{2}-\lambda^{2}}\cos\left[\sqrt{\alpha^{2}-\lambda^{2}}(L-a)\right]
+Ua_{I}\sin\left(\sqrt{\alpha^{2}-\lambda^{2}}\,a\right)\;,\;\;\;\;\;
\end{eqnarray}
which has a non-trivial solution for the coefficients $a_{I}$ and $a_{II}$ if the determinant of the coefficient matrix vanishes identically,
namely
\begin{equation}\label{15}
\Omega_{\lambda}^{DD}(\alpha,a)=\sqrt{\alpha^{2}-\lambda^{2}}\sin\left(\sqrt{\alpha^{2}-\lambda^{2}}\,L\right)-U\sin\left(\sqrt{\alpha^{2}-\lambda^{2}}\,a\right)\sin\left[\sqrt{\alpha^{2}-\lambda^{2}}(L-a)\right]=0\;.
\end{equation}
The above equation implicitly determines the eigenvalues $\alpha$ for the Dirichlet-Dirichlet case. For the case of Neumann-Neumann boundary conditions
we obtain $a_{I}=a_{II}=0$ and the system
\begin{eqnarray}\label{16}
b_{I}\cos\left(\sqrt{\alpha^{2}-\lambda^{2}}\,a\right)&=&b_{II}\cos\left[\sqrt{\alpha^{2}-\lambda^{2}}(L-a)\right]\nonumber\\
-b_{I}\sqrt{\alpha^{2}-\lambda^{2}}\sin\left(\sqrt{\alpha^{2}-\lambda^{2}}\,a\right)&=&b_{II}\sqrt{\alpha^{2}-\lambda^{2}}\sin\left[\sqrt{\alpha^{2}-\lambda^{2}}(L-a)\right]
+Ub_{I}\cos\left(\sqrt{\alpha^{2}-\lambda^{2}}\,a\right)\;,\;\;\;\;\;
\end{eqnarray}
which one can get by imposing transmittal boundary conditions at $N_{a}$.
The system (\ref{16}) has a non-trivial solution for $b_{I}$ and $b_{II}$ if
\begin{equation}\label{17}
\Omega_{\lambda}^{NN}(\alpha,a)=\sqrt{\alpha^{2}-\lambda^{2}}\sin\left(\sqrt{\alpha^{2}-\lambda^{2}}\,L\right)-U\cos\left(\sqrt{\alpha^{2}-\lambda^{2}}\,a\right)\cos\left[\sqrt{\alpha^{2}-\lambda^{2}}(L-a)\right]=0\;,
\end{equation}
which is the equation determining the eigenvalues $\alpha$ in the Neumann-Neumann case. The exact same procedure can be followed to obtain an implicit equation for the
eigenvalues in the case of mixed boundary conditions. In fact, by imposing mixed boundary conditions of the Dirichlet-Neumann type at the endpoints of $M$ and
transmittal boundary conditions at $N_{a}$ and by subsequently setting to zero the determinant of the coefficient matrix of the ensuing linear system we obtain
\begin{equation}\label{18}
\Omega_{\lambda}^{DN}(\alpha,a)=\sqrt{\alpha^{2}-\lambda^{2}}\cos\left(\sqrt{\alpha^{2}-\lambda^{2}}\,L\right)-U\sin\left(\sqrt{\alpha^{2}-\lambda^{2}}\,a\right)\cos\left[\sqrt{\alpha^{2}-\lambda^{2}}(L-a)\right]=0\;.
\end{equation}
For the other type of mixed boundary conditions, namely Neumann-Dirichlet, we obtain, instead, the following equation for the eigenvalues $\alpha$
\begin{equation}\label{19}
\Omega_{\lambda}^{ND}(\alpha,a)=\sqrt{\alpha^{2}-\lambda^{2}}\cos\left(\sqrt{\alpha^{2}-\lambda^{2}}\,L\right)-U\cos\left(\sqrt{\alpha^{2}-\lambda^{2}}\,a\right)\sin\left[\sqrt{\alpha^{2}-\lambda^{2}}(L-a)\right]=0\;.
\end{equation}

The equations (\ref{15}), and (\ref{17})-(\ref{19}) can be used to write an expression for the spectral zeta function in terms of a contour integral,
valid in the semi-plane $\Re(s)>D/2$, as follows \cite{bordag96a,bordag96b,kirsten01}
\begin{equation}\label{20}
\zeta^{(j)}(s,a)=\frac{1}{2\pi i}\sum_{\lambda}d(\lambda)\int_{\gamma_{j}}\kappa^{-2s}\frac{\partial}{\partial\kappa}\ln\Omega^{(j)}_{\lambda}(\kappa,a)\,\diff\kappa\;,
\end{equation}
where the index $j$ denotes the type of boundary conditions under consideration and $\gamma_{j}$ represents a contour that encloses, in the counterclockwise direction, all the real zeroes of the appropriate implicit equation for the eigenvalues $\Omega^{(j)}_{\lambda}(\kappa,a)=0$. In addition, $d(\lambda)$ indicates the degeneracy of the
eigenvalues $\lambda$ of the Laplacian on the manifold $N$.
According to (\ref{4}) we need to analyze the spectral zeta function in a neighborhood of $s=-1/2$ in order to obtain information about the Casimir energy and force. Since the point $s=-1/2$ falls outside the region of validity of the
integral representation (\ref{20}) an analytic continuation to the complementary region $\Re(s)\leq D/2$ must be performed.

\section{Analytic continuation of the spectral zeta function}\label{sec3}

The first step of the desired analytic continuation is performed by exploiting the replacement $\kappa\to z\lambda$ and by deforming the integration contour $\gamma_{j}$
in (\ref{20}) to the imaginary axis \cite{kirsten01}.
This procedure allows us to rewrite the spectral zeta function as
\begin{eqnarray}\label{21}
\zeta^{(j)}(s,a)=\sum_{\lambda}d(\lambda)\zeta_{\lambda}^{(j)}(s,a)\;,
\end{eqnarray}
where $\zeta_{\lambda}^{(j)}(s,a)$ is represented as a real integral of the form
\begin{equation}\label{22}
\zeta_{\lambda}^{(j)}(s,a)=\frac{\sin(\pi s)}{\pi}\lambda^{-2s}\int_{0}^{\infty}z^{-2s}\frac{\partial}{\partial z}\ln\Omega^{(j)}_{\lambda}(i\lambda z,a)\diff z\;.
\end{equation}
The functions $\Omega^{(j)}_{\lambda}(i\lambda z,a)$ can be obtained from (\ref{15}), and (\ref{17})-(\ref{19}) and explicitly read
\begin{eqnarray}\label{23}
\Omega_{\lambda}^{DD}(i\lambda z,a)&=&\lambda\sqrt{1+z^{2}}\sinh\left(\lambda\sqrt{1+z^{2}}\,L\right)-U\sinh\left(\lambda\sqrt{1+z^{2}}\,a\right)\sinh\left[\lambda\sqrt{1+z^{2}}(L-a)\right]\;,\label{23a}\\
\Omega_{\lambda}^{NN}(i\lambda z,a)&=&\lambda\sqrt{1+z^{2}}\sinh\left(\lambda\sqrt{1+z^{2}}\,L\right)-U\cosh\left(\lambda\sqrt{1+z^{2}}\,a\right)\cosh\left[\lambda\sqrt{1+z^{2}}(L-a)\right]\;,\label{23b}\\
\Omega_{\lambda}^{DN}(i\lambda z,a)&=&\lambda\sqrt{1+z^{2}}\cosh\left(\lambda\sqrt{1+z^{2}}\,L\right)-U\sinh\left(\lambda\sqrt{1+z^{2}}\,a\right)\cosh\left[\lambda\sqrt{1+z^{2}}(L-a)\right]\;,\label{23c}\\
\Omega_{\lambda}^{ND}(i\lambda z,a)&=&\lambda\sqrt{1+z^{2}}\cosh\left(\lambda\sqrt{1+z^{2}}\,L\right)-U\cosh\left(\lambda\sqrt{1+z^{2}}\,a\right)\sinh\left[\lambda\sqrt{1+z^{2}}(L-a)\right]\;.\label{23d}
\end{eqnarray}
In order for the contour deformation to be well defined, one needs to make sure that no zeroes of (\ref{15}), and (\ref{17})-(\ref{19}) lie on the imaginary axis.
However, it is not difficult to realize that the solutions to the implicit equations for $\alpha$ in (\ref{15}), and (\ref{17})-(\ref{19}) are simple and can
be either real or imaginary \cite{fucci15,romeo02,teo09}.
Since we only want to consider boundary conditions that lead to a self-adjoint boundary value problem, we restrict our analysis to values of the parameter $U$ for
which the zeroes of (\ref{15}), and (\ref{17})-(\ref{19}) are real and positive. This assumption also allows the contour deformation leading to (\ref{22}) to be well defined.
A discussion of the case in which purely imaginary zeroes are present can be found in \cite{romeo02}. To find the range of allowed values of $U$, we start by noticing that
the equations (\ref{15}), and (\ref{17})-(\ref{19}) have purely imaginary zeroes if (\ref{23a})-(\ref{23d}) have real zeroes. In particular, if we denote by $\lambda_{0}>0$ the smallest eigenvalue of $-\Delta_{N}$, then (\ref{23a}) has no real zeroes if
\begin{equation}\label{24}
  \lambda_{0}\sinh\left(\lambda_{0}L\right)>U\sinh\left(\lambda_{0}a\right)\sinh\left[\lambda_{0}(L-a)\right]\;.
\end{equation}
The last inequality holds for all $a\in(0,L)$ if $U<U_{DD}$ where
\begin{equation}\label{25}
  U_{DD}=\frac{\lambda_{0}\sinh\left(\lambda_{0}L\right)}{\left[\sinh\left(\frac{\lambda_{0}L}{2}\right)\right]^{2}}\;.
\end{equation}
Similarly, the function in (\ref{23b}) has no real zeroes if $U<U_{NN}$ with
\begin{equation}\label{26}
   U_{NN}=\lambda_{0}\tanh\left(\lambda_{0}L\right)\;.
\end{equation}
Lastly, the functions describing mixed boundary conditions (\ref{23c}) and (\ref{23d}) have no real zeroes for values of $U<U_{M}$ where one finds that
\begin{equation}\label{27}
  U_{M}=\frac{\lambda_{0}}{\tanh\left(\lambda_{0}L\right)}\;.
\end{equation}
For the majority of compact Riemannian manifolds $N$, and appropriate boundary conditions, the lowest eigenvalue $\lambda_{0}$ cannot be explicitly found.
However, lower bounds for the smallest eigenvalue of the Laplacian can be found (see e.g. \cite{cheeger,chavel}). Such lower bounds can then be utilized in 
the formulas (\ref{25})-(\ref{27}) to obtain an upper bound on the allowed values of $U$.
We would like to point out that if $\lambda_{0}=0$ then the above analysis need to be slightly amended. 
Later in this work we will outline a method that can be used to analytically continue the spectral zeta function when $\Delta_{N}$ has a zero mode.
We can, finally, conclude that when $U<U_{(j)}$ then the contour deformation can be performed without any problems and leads to the formulas (\ref{21}) and (\ref{22}).
By analyzing the behavior of $z^{-2s}\partial_{z}\ln\Omega^{(j)}_{\lambda}(i\lambda z,a)$ for $z\to 0$ and $z\to\infty$ it is not very difficult to show (see e.g. \cite{fucci15})
that the integral representation (\ref{22}) is valid in the strip $1/2<\Re(s)<1$.
In order to extend the region of convergence of the integral (\ref{22}) to $\Re(s)\leq1/2$, we
subtract, and add, in the representation (\ref{22}), a suitable number of terms of the asymptotic expansion of $\ln\Omega^{(j)}_{\lambda}(i\lambda z,a)$ for $\lambda\to\infty$
uniform in $z=k/\lambda$.

By using the explicit expressions (\ref{23a}) through (\ref{23d}) and the exponential form of the hyperbolic functions, one can prove that
\begin{equation}\label{28}
  \Omega_{\lambda}^{(j)}(i\lambda z,a)=\frac{1}{2}e^{\lambda\sqrt{1+z^{2}}L}\left(\lambda\sqrt{1+z^{2}}-\frac{U}{2}\right)\left[1+\textrm{exp}_{(j)}(z,\lambda,a)\right]\;,
\end{equation}
where $\textrm{exp}_{(j)}(z,\lambda,a)$ represents exponentially small terms as $\lambda\to\infty$. From the expression (\ref{28}) it is not difficult to obtain the following one
\begin{equation}\label{29}
  \ln\Omega_{\lambda}^{(j)}(i\lambda z,a)=\lambda\sqrt{1+z^{2}}L-\ln 2+\ln\left(\lambda\sqrt{1+z^{2}}\right)+\ln\left(1-\frac{U}{2\lambda\sqrt{1+z^{2}}}\right)+\ln\left[1+\textrm{exp}_{(j)}(z,\lambda,a)\right]\;.
\end{equation}
By exploiting the small-$x$ asymptotic expansion of $\ln(1-x)$ we find the needed uniform asymptotic expansion
\begin{equation}\label{30}
  \ln\Omega_{\lambda}^{(j)}(i\lambda z,a)\sim\lambda\sqrt{1+z^{2}}L-\ln 2+\ln\left(\lambda\sqrt{1+z^{2}}\right)-\sum_{n=1}^{\infty}\frac{U^{n}}{2^{n}n}\frac{1}{\lambda^{n}(1+z^{2})^{\frac{n}{2}}}\;,
\end{equation}
where we have discarded the exponentially small terms.

By subtracting and adding $N$ terms of the uniform asymptotic expansion (\ref{30}) in the integral representation (\ref{22})
we obtain, according to (\ref{21}), the following expression
\begin{equation}\label{31}
\zeta^{(j)}(s,a)=Z^{(j)}(s,a)+\sum_{k=-1}^{N}A_{k}(s,a)\;.
\end{equation}
The function $Z^{(j)}(s,a)$ is analytic in the region of the complex plane $\Re(s)>(d-N-1)/2$ and takes the form
\begin{eqnarray}\label{32}
 Z^{(j)}(s,a)&=&\frac{\sin(\pi s)}{\pi}\sum_{\lambda}d(\lambda)\lambda^{-2s}\int_{0}^{\infty}z^{-2s}\frac{\partial}{\partial z}\Bigg[\ln\Omega^{(j)}_{\lambda}(i\lambda z,a)-\lambda\sqrt{1+z^{2}}L+\ln 2\nonumber\\
 &-&\ln\left(\lambda\sqrt{1+z^{2}}\right)+\sum_{n=1}^{N}\frac{U^{n}}{2^{n}n}\frac{1}{\lambda^{n}(1+z^{2})^{\frac{n}{2}}}\Bigg]\diff z\;.
\end{eqnarray}
The terms $A_{k}(s,a)$ are, instead, meromorphic functions of $s$ in the entire complex plane
and can be written as
\begin{equation}\label{33}
  A_{-1}(s,a)=\frac{\sin(\pi s)}{\pi}\sum_{\lambda}d(\lambda)\lambda^{-2s}\int_{0}^{\infty}z^{-2s}\frac{\partial}{\partial z}\left(\lambda\sqrt{1+z^{2}}L\right)\diff z\;,
\end{equation}
\begin{equation}\label{34}
  A_{0}(s,a)=\frac{\sin(\pi s)}{\pi}\sum_{\lambda}d(\lambda)\lambda^{-2s}\int_{0}^{\infty}z^{-2s}\frac{\partial}{\partial z}\left[\ln\left(\lambda\sqrt{1+z^{2}}\right)\right]\diff z\;,
\end{equation}
and, for $k\geq 1$,
\begin{equation}\label{35}
   A_{k}(s,a)=\frac{\sin(\pi s)}{\pi}\sum_{\lambda}d(\lambda)\lambda^{-2s-k}\frac{U^{k}}{2^{k}k}\int_{0}^{\infty}z^{-2s}\frac{\partial}{\partial z}\left[(1+z^{2})^{-\frac{k}{2}}\right]\diff z\;.
\end{equation}
By performing the elementary integrals (\ref{33})-(\ref{35}) and by using the following definition for the spectral zeta function
associated with the Laplacian $\Delta_{N}$
\begin{equation}\label{31a}
  \zeta_{N}(s)=\sum_{\lambda}d(\lambda)\lambda^{-2s}\;,
\end{equation}
we can finally write the analytically continued expression for the spectral zeta function as
\begin{eqnarray}\label{36}
  \zeta^{(j)}(s,a)&=&Z^{(j)}(s,a)+\frac{L}{2\sqrt{\pi}\Gamma(s)}\Gamma\left(s-\frac{1}{2}\right)\zeta_{N}\left(s-\frac{1}{2}\right)+\frac{1}{2}\zeta_{N}(s)\nonumber\\
  &-&\frac{1}{\Gamma(s)}\sum_{k=1}^{N}\frac{U^{k}}{2^{k+1}\Gamma\left(\frac{k}{2}+1\right)}\Gamma\left(s+\frac{k}{2}\right)\zeta_{N}\left(s+\frac{k}{2}\right)\;.
\end{eqnarray}
As it is rendered manifest in the above expression, the meromorphic structure of the spectral zeta function is completely determined by the terms in (\ref{36})
proportional to the zeta function $\zeta_{N}(s)$.

\section{Presence of zero modes }\label{sec4}

The analytic continuation outlined in the previous sections holds if $\Delta_{N}$ has non-vanishing eigenvalues. If, instead, $\lambda=0$ is an eigenvalue 
of $\Delta_{N}$ with multiplicity $d(0)$, then the process of analytic continuation of the spectral zeta function $\zeta(s,a)$ needs to be somewhat amended since we can 
no longer utilize the uniform asymptotic expansion in (\ref{30}). When a zero mode is present, it is convenient to separate its contribution to the spectral zeta function from the one of 
the non-vanishing eigenvalues as follows   
\begin{equation}\label{37}
\zeta^{(j)}(s,a)=\frac{d(0)}{2\pi i}\int_{\gamma_{j}}\kappa^{-2s}\frac{\partial}{\partial\kappa}\ln\Omega^{(j)}_{0}(\kappa,a)\,\diff\kappa+
\sum_{\lambda>0}\frac{d(\lambda)}{2\pi i}\int_{\gamma_{j}}\kappa^{-2s}\frac{\partial}{\partial\kappa}\ln\Omega^{(j)}_{\lambda}(\kappa,a)\,\diff\kappa\;.
\end{equation}
The analytic continuation of the contribution coming from the non-vanishing eigenvalues follows exactly the same process described earlier and, therefore, 
will not be repeated here. We direct, instead, our attention to the analytic continuation of the first integral on the right-hand-side of (\ref{37}). When $\lambda=0$ is an eigenvalue, 
the differential equation (\ref{2}) becomes 
\begin{equation}\label{38}
  \left(-\frac{\diff^2}{\diff x^2}-\alpha^{2}\right)f_{\lambda}(x,\alpha)=0\;,
\end{equation}
whose general solution in chamber $I$ and chamber $II$ is simply $f_{I,0}(x,\alpha)$ from (\ref{8}) and $f_{II,0}(x,\alpha)$ from (\ref{9}), respectively.
To these solutions we need to impose appropriate boundary conditions. By applying Dirichlet-Dirichlet boundary conditions with transmittal ones on the piston itself,
we obtain the following equation for the eigenvalues $\alpha$
\begin{equation}\label{39}
\Omega^{DD}_{0}(\alpha,a)=0\;.
\end{equation}
The equations that determine the eigenvalues when the other boundary conditions are imposed, namely Neumann-Neumann and mixed, are found to be
\begin{equation}\label{40}
 \Omega^{NN}_{0}(\alpha,a)=0\;,\quad \Omega^{DN}_{0}(\alpha,a)=0\;, \quad \textrm{and}\quad \Omega^{ND}_{0}(\alpha,a)=0\;. 
\end{equation}

By performing the replacement $\alpha\to iz$ and by deforming the contour to the imaginary axis we obtain, for the zero 
mode contribution to the spectral zeta function, the integral representation
\begin{equation}\label{41}
\zeta_{0}^{(j)}(s,a)=d(0)\frac{\sin(\pi s)}{\pi}\lambda^{-2s}\int_{0}^{\infty}z^{-2s}\frac{\partial}{\partial z}\ln\Omega^{(j)}_{0}(i z,a)\diff z\;,
\end{equation}
which, just like earlier, is well defined in the strip $1/2<\Re(s)<1$ and where the functions $\Omega^{(j)}_{0}(i z,a)$ can be shown to have the form
\begin{eqnarray}
\Omega_{0}^{DD}(i z,a)&=&z\sinh\left(z L\right)-U\sinh\left(z a\right)\sinh\left[z(L-a)\right]\;,\label{42a}\\
\Omega_{0}^{NN}(i z,a)&=&z\sinh\left(z L\right)-U\cosh\left(z a\right)\cosh\left[z(L-a)\right]\;,\label{42b}\\
\Omega_{0}^{DN}(i z,a)&=&z\cosh\left(z L\right)-U\sinh\left(z a\right)\cosh\left[z(L-a)\right]\;,\label{42c}\\
\Omega_{0}^{ND}(i z,a)&=&z\cosh\left(z L\right)-U\cosh\left(z a\right)\sinh\left[z(L-a)\right]\;.\label{42d}
\end{eqnarray} 
Now, the equations (\ref{39}) and (\ref{40}) have, in principle, both real and imaginary solutions. For reasons already explained in the previous sections,
we need to find the rage of values of the parameter $U$ for which (\ref{39}) and (\ref{40}) have only real solutions. This is achieved for values of 
$U$ for which (\ref{42a}) through (\ref{42d}) have no real solutions. The allowed values of $U$ satisfy the inequality $U<U_{0,(j)}$ where 
the quantity $U_{0,(J)}$ can be obtained from the equations (\ref{25})-(\ref{27}) by taking the limit $\lambda_{0}\to 0$. In more details, they read,
\begin{equation}\label{43}
  U_{0,DD}=\frac{4}{L}\;,\quad  U_{0,NN}=0\;,\quad \textrm{and}\quad U_{0,M}=\frac{1}{L}\;.
\end{equation}    
Hence, when $U<U_{0,(j)}$, the contour deformation can be performed and leads to the representation (\ref{41}) for the zero mode contribution to the 
spectral zeta function. The process of analytic continuation begins by splitting the integral representation (\ref{41}) \cite{fucci15} as
\begin{equation}\label{44}
\zeta_{0}^{(j)}(s,a)=d(0)\frac{\sin(\pi s)}{\pi}\lambda^{-2s}\int_{0}^{1}z^{-2s}\frac{\partial}{\partial z}\ln\Omega^{(j)}_{0}(i z,a)\diff z
+d(0)\frac{\sin(\pi s)}{\pi}\lambda^{-2s}\int_{1}^{\infty}z^{-2s}\frac{\partial}{\partial z}\ln\Omega^{(j)}_{0}(i z,a)\diff z\;.
\end{equation}
The advantage of rewriting (\ref{41}) as the sum in (\ref{44}) lies in the fact that the first integral is analytic in the semi-plane $\Re(s)<1$, and hence no further manipulation is necessary for the purpose of analyzing $\zeta_{0}^{(j)}(s,a)$ at $s=-1/2$. The second integral is analytic for $\Re(s)>1/2$, and therefore, requires to be extended to 
the left of the abscissa of convergence $\Re(s)=1/2$. To perform the analytic continuation of the second integral in (\ref{44}) we proceed as before by subtracting and adding a suitable number of terms of the large-$z$ asymptotic expansion of $\ln\Omega^{(j)}_{0}(i z,a)$ from the second integral in (\ref{44}). From the expressions (\ref{42a})-(\ref{42d})
it is not very difficult to find the relation 
\begin{equation}\label{45}
  \ln\Omega^{(j)}_{0}(i z,a)=zL-\ln 2+\ln z+\ln\left(1-\frac{U}{2z}\right)+\ln\left[1+\textrm{exp}_{0,(j)}(z,\lambda,a)\right]\;,
\end{equation} 
where $\textrm{exp}_{0,(j)}(z,\lambda,a)$ denotes exponentially decaying terms. The desired large-$z$ asymptotic expansion is obtained by expanding 
$\ln(1-U/2z)$ for small values of $U/2z$ and by discarding exponentially small terms in (\ref{45}). This procedure leads to 
\begin{equation}\label{46}
   \ln\Omega^{(j)}_{0}(i z,a)\sim zL-\ln 2+\ln z+\sum_{n=1}^{\infty}\frac{U^{n}}{2^{n}n}\frac{1}{z^{n}}\;.
\end{equation}

By subtracting and adding from the second integral in (\ref{44}), $N$ terms of the asymptotic expansion (\ref{46}) and by then performing the remaining trivial integrals we obtain, for the zero mode contribution to the spectral zeta function, the expression
\begin{equation}\label{47}
  \zeta_{0}^{(j)}(s,a)=Z_{0}^{(j)}(s,a)+d(0)\frac{\sin(\pi s)}{\pi}\left[\frac{L}{2s-1}+\frac{1}{2s}-\sum_{k=1}^{N}\left(\frac{U}{2}\right)^{k}\frac{1}{2s+k}\right]\;,
\end{equation} 
where $Z_{0}^{(j)}(s,a)$ is an analytic function for $\Re(s)>-(N+1)/2$ having the integral representation
\begin{eqnarray}\label{48}
  Z_{0}^{(j)}(s,a)=d(0)\frac{\sin(\pi s)}{\pi}\int_{0}^{\infty}z^{-2s}\frac{\partial}{\partial z}\left\{\ln\Omega^{(j)}_{0}(i z,a)-\Theta(z-1)\left[zL-\ln 2+\ln z+\sum_{n=1}^{N}\frac{U^{n}}{2^{n}n}\frac{1}{z^{n}}\right]\right\}\;,
\end{eqnarray} 
with $\Theta(z)$ denoting the Heaviside step-function. Once again, the meromorphic structure of $\zeta_{0}^{(j)}(s,a)$ is completely encoded in the terms in square parentheses in
(\ref{47}).

\section{Computation of the Casimir energy and force}\label{sec5}

The analytic continuation of the spectral zeta function $\zeta^{(j)}(s,a)$ in (\ref{36}) can be used, at this point, to compute the Casimir energy of the 
system. According to the definition given in (\ref{4}) we need to analyze the structure of $\zeta^{(j)}(s,a)$ in a neighborhood of $s=-1/2$. In order to obtain 
an expression valid in a neighborhood of $s=-1/2$ it is sufficient to set $N=D$ in the analytic continuation (\ref{36}), namely
\begin{eqnarray}\label{49}
  \zeta^{(j)}(s,a)&=&Z^{(j)}(s,a)+\frac{L}{2\sqrt{\pi}\Gamma(s)}\Gamma\left(s-\frac{1}{2}\right)\zeta_{N}\left(s-\frac{1}{2}\right)+\frac{1}{2}\zeta_{N}(s)\nonumber\\
  &-&\frac{1}{\Gamma(s)}\sum_{k=1}^{D}\frac{U^{k}}{2^{k+1}\Gamma\left(\frac{k}{2}+1\right)}\Gamma\left(s+\frac{k}{2}\right)\zeta_{N}\left(s+\frac{k}{2}\right)\;.
\end{eqnarray}
The representation (\ref{49}) is now well defined for $\Re(s)>-1$. To compute $\zeta^{(j)}(s,a)$ at $s=-1/2$ it is convenient to
make, in (\ref{49}), the substitution $s=\epsilon-1/2$ and then analyze the ensuing small-$\epsilon$ expansion. 

Since $Z^{(j)}(s,a)$ is analytic for $\Re(s)>-1$ we can simply set, in its expression, $s=-1/2$.
To perform the small-$\epsilon$ expansion for the remaining terms on the right-hand-side of (\ref{49}) we need to take into account the meromorphic structure 
of the spectral zeta function $\zeta_{N}(s)$ which, according to the general theory \cite{gilkey95,kirsten01}, is 
\begin{eqnarray}
\zeta_{N}(\epsilon -n)&=&\zeta_{N}(-n)+\epsilon\zeta^{\prime}_{N}(-n)+O(\epsilon^{2})\;,\label{50}\\
\zeta_{N}\left(\epsilon+\frac{d-k}{2}\right)&=&\frac{1}{\epsilon}\textrm{Res}\,\zeta_{N}\left(\frac{d-k}{2}\right)+\textrm{FP}\,\zeta_{N}\left(\frac{d-k}{2}\right)+O(\epsilon)\;,\label{50a}\\
\zeta_{N}\left(\epsilon-\frac{2n+1}{2}\right)&=&\frac{1}{\epsilon}\textrm{Res}\,\zeta_{N}\left(-\frac{2n+1}{2}\right)+\textrm{FP}\,\zeta_{N}\left(-\frac{2n+1}{2}\right)+O(\epsilon)\;,\label{50b}
\end{eqnarray}
where $n\in\mathbb{N}_{0}$ and $k=\{0,\ldots,d-1\}$. It is important, at this point, to mention that the residues of the spectral zeta function $\zeta_{N}(s)$
are proportional to the coefficient of the small-$t$ asymptotic expansion of the trace of the heat kernel associated with the Laplace operator $\Delta_{N}$ \cite{gilkey95,gilkey04,kirsten01}, that is
\begin{equation}\label{51}
  \Gamma\left(\frac{d-k}{2}\right)\textrm{Res}\,\zeta_{N}\left(\frac{d-k}{2}\right)=A^{N}_{\frac{k}{2}}\;,\quad
  \Gamma\left(-\frac{2n+1}{2}\right)\textrm{Res}\,\zeta_{N}\left(-\frac{2n+1}{2}\right)=A^{N}_{\frac{d+2n+1}{2}}\;.
\end{equation}

By substituting $s=\epsilon-1/2$ in 
the second term on the right-hand-side of (\ref{49}) and by using the relation (\ref{50}) we obtain
\begin{equation}\label{52}
  \frac{L}{2\sqrt{\pi}\Gamma\left(\epsilon-\frac{1}{2}\right)}\Gamma(\epsilon-1)\zeta_{N}(\epsilon-1)=\frac{L\,\zeta_{N}(-1)}{4\pi\epsilon}
  +\frac{L}{4\pi}\left[\zeta'_{N}(-1)+(2\ln 2-1)\zeta_{N}(-1)\right]+O(\epsilon)\;.
\end{equation} 
For the next term, we utilize the expression (\ref{50b}) with $n=0$ to find the expansion
\begin{equation}\label{53}
\frac{1}{2}\zeta_{N}\left(\epsilon-\frac{1}{2}\right)=\frac{1}{2\epsilon}\textrm{Res}\,\zeta_{N}\left(-\frac{1}{2}\right)+\frac{1}{2}\textrm{FP}\,\zeta_{N}\left(-\frac{1}{2}\right)
+O(\epsilon)\;.
\end{equation}
In the sum appearing in (\ref{49}) we need to separate the contribution of the term with $k=1$ from the rest. When $k=1$ we have the expansion
\begin{equation}\label{54}
-\frac{U}{2\sqrt{\pi}\Gamma\left(\epsilon-\frac{1}{2}\right)}\Gamma(\epsilon)\zeta_{N}(\epsilon)=\frac{U}{4\pi\epsilon}\zeta_{N}(0)+
\frac{U}{4\pi}\left[\zeta'_{N}(0)+2(\ln 2-1)\zeta_{N}(0)\right]+O(\epsilon)\;,
\end{equation} 
which can be obtained thanks to (\ref{50}). For the terms of the sum with $k=\{2,\ldots,D\}$ we exploit (\ref{50a}) to get
\begin{eqnarray}\label{55}
&&\lefteqn{-\frac{U^{k}}{2^{k+1}\Gamma\left(\frac{k}{2}+1\right)}\frac{\Gamma\left(\epsilon+\frac{k-1}{2}\right)}{\Gamma\left(\epsilon-\frac{1}{2}\right)}\zeta_{N}\left(\epsilon+\frac{k-1}{2}\right)=\frac{U^{k}}{2^{k+2}\sqrt{\pi}\Gamma\left(\frac{k}{2}+1\right)\epsilon}\Gamma\left(\frac{k-1}{2}\right)\textrm{Res}\,\zeta_{N}\left(\frac{k-1}{2}\right)}\nonumber\\
&+&\frac{U^{k}}{2^{k+2}\sqrt{\pi}\Gamma\left(\frac{k}{2}+1\right)}\Gamma\left(\frac{k-1}{2}\right)\left\{\textrm{FP}\,\zeta_{N}\left(\frac{k-1}{2}\right)
+\left[2-\gamma-2\ln 2+\Psi\left(\frac{k-1}{2}\right)\right]\textrm{Res}\,\zeta_{N}\left(\frac{k-1}{2}\right)\right\}\;.
\end{eqnarray}

The results obtained above allow us to write the expression for the Casimir energy of the piston as follows
\begin{eqnarray}\label{56}
E_{\textrm{Cas}}^{(j)}(a)&=&\frac{1}{2}\left(\frac{1}{\epsilon}+\ln\mu^2\right)\Bigg[\frac{L}{4\pi}\zeta_{N}(-1)+\frac{1}{2}\textrm{Res}\,\zeta_{N}\left(-\frac{1}{2}\right)
+\frac{U}{4\pi}\zeta_{N}(0)\nonumber\\
&+&\sum_{k=2}^{D}\frac{U^{k}}{2^{k+2}\sqrt{\pi}\Gamma\left(\frac{k}{2}+1\right)}\Gamma\left(\frac{k-1}{2}\right)\textrm{Res}\,\zeta_{N}\left(\frac{k-1}{2}\right)\Bigg]
+\frac{1}{2}Z^{(j)}\left(-\frac{1}{2},a\right)\nonumber\\
&+&\frac{L}{4\pi}\left[\zeta'_{N}(-1)+(2\ln 2-1)\zeta_{N}(-1)\right]+\frac{1}{2}\textrm{FP}\,\zeta_{N}\left(-\frac{1}{2}\right)
+\frac{U}{4\pi}\left[\zeta'_{N}(0)+2(\ln 2-1)\zeta_{N}(0)\right]\nonumber\\
&+&\sum_{k=2}^{D}\frac{U^{k}\Gamma\left(\frac{k-1}{2}\right)}{2^{k+2}\sqrt{\pi}\Gamma\left(\frac{k}{2}+1\right)}\left\{\textrm{FP}\,\zeta_{N}\left(\frac{k-1}{2}\right)
+\left[2-\gamma-2\ln 2+\Psi\left(\frac{k-1}{2}\right)\right]\textrm{Res}\,\zeta_{N}\left(\frac{k-1}{2}\right)\right\}\nonumber\\
&+&O(\epsilon)\;.
\end{eqnarray} 
It is clear from this expression that the Casimir force for the piston configuration is, in general, not a well-defined quantity as has already been observed before \cite{bordag09}. The ambiguity in the energy is proportional to the coefficients, which encode geometric information of the manifold $N$, 
of the asymptotic expansion of the heat kernel associated with the Laplacian on $N$ as one can easily infer from the relations (\ref{51}) and the following one \cite{gilkey95,kirsten01}
\begin{equation}
A^{N}_{\frac{d}{2}+p}=(-1)^{p}\frac{\zeta_{N}(-p)}{p!}\;,
\end{equation}  
with $p\in\mathbb{N}_{0}$. To complete the result about the Casimir energy we would like to consider the contribution to it coming from 
possible zero modes of the Laplacian on $N$. From the expression (\ref{47}) for $\zeta_{0}^{(j)}(s,a)$, we set $N=1$, $s=\epsilon-1/2$, and compute the small-$\epsilon$ 
expansion of the resulting formula to obtain 
\begin{equation}\label{57}
E_{\textrm{Cas},0}^{(j)}(a)=\frac{d(0)}{8\pi}\left(\frac{1}{\epsilon}+\ln\mu^2\right)U+\frac{1}{2}Z^{(j)}_{0}\left(-\frac{1}{2},a\right)
+\frac{d(0)}{4\pi}(L+2)\;,
\end{equation} 
which, just like before, is in general not a well-defined quantity.

Although the Casimir energy is, in general, an ambiguous quantity, the force acting on the piston is well defined. In fact, it is easy to show, according to the definition (\ref{6}), 
that the Casimir force acting on the piston is simply
\begin{equation}\label{58}
F^{(j)}_{\textrm{Cas}}(a)=-\frac{1}{2}\left(Z^{(j)}\right)'\left(-\frac{1}{2},a\right)\;.
\end{equation} 
Analogously, if the Laplacian on $N$ has zero modes then their contribution to the Casimir force on the piston can be obtained from (\ref{6}) and (\ref{57}), 
more explicitly  
\begin{equation}\label{59}
F^{(j)}_{\textrm{Cas},0}(a)=-\frac{1}{2}\left(Z^{(j)}_{0}\right)'\left(-\frac{1}{2},a\right)\;.
\end{equation}

\section{A $d$-dimensional sphere as piston}\label{sec6}

In this section we apply the results for the Casimir force obtained earlier to the case of a piston configuration where 
the piston itself is assumed to be a unit $d$-dimensional sphere. For simplicity we also assume that the length of the piston configuration is $L=1$.  
For a $d$-dimensional sphere, the eigenvalues of the Laplacian on $N$ are explicitly 
known and can be written as
\begin{equation}\label{79}
\lambda=l+\frac{d-1}{2}\;,
\end{equation}  
where $l\in\mathbb{N}_{0}$. The eigenfunctions on $N$ are found to be hyperspherical harmonics with degeneracy 
\begin{equation}\label{80}
d(l)=(2l+d-1)\frac{(l+d-2)!}{l!(d-1)!}\;.
\end{equation}
By using (\ref{79}) and (\ref{80}) the spectral zeta function on the manifold $N$ can be expressed as
\begin{equation}\label{81}
\zeta_{N}(s)=\sum_{l=0}^{\infty}(2l+d-1)\frac{(l+d-2)!}{l!(d-1)!}\left(l+\frac{d-1}{2}\right)^{-2s}\;,
\end{equation}
which, in turn, can be written as a linear combination of Hurwitz zeta functions \cite{bordag96,bordag96a,fucci11,fucci11b}
\begin{equation}\label{82}
\zeta_{N}(s)=2\sum_{\alpha=0}^{d-1}e_{\alpha}\zeta_{H}\left(2s-\alpha-1,\frac{d-1}{2}\right)\;,
\end{equation}
with the coefficients $e_{\alpha}$ determined according to the formula
\begin{equation}\label{83}
\frac{(l+d-2)!}{l!(d-1)!}=\sum_{\alpha=0}^{d-1}e_{\alpha}\left(l+\frac{d-1}{2}\right)^{\alpha}\;.
\end{equation}
By using the eigenvalues (\ref{79}) and their degeneracy (\ref{80}) we can now analyze explicitly the Casimir force on the piston (\ref{58}) in the cases of
Dirichlet-Dirichlet, Neumann-Neumann, and mixed boundary conditions. In what follows we set, for definiteness, $d=2$. Obviously the analysis can be easily carried out 
in any dimension $d$. 

\subsection{Dirichlet-Dirichlet boundary conditions}

\begin{figure}[]
\centering
\includegraphics[scale=0.60,trim=0cm 0cm 0cm 0cm, clip=true, angle=0]{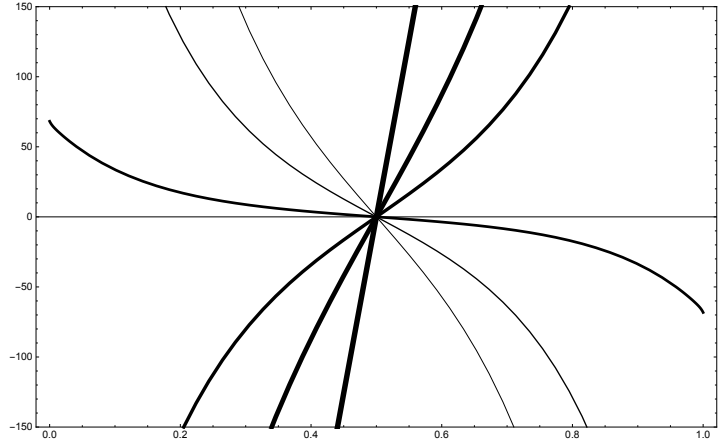}
\caption{Dirichlet boundary conditions at $x=0$ and $x=1$, and transmittal boundary conditions at $x=a$. The values along the $y$-axis provide the magnitude (in units for which $h=c=1$) of the Casimir force on the piston.}\label{fig1}
\end{figure}

First, we consider the case in which Dirichlet boundary conditions are imposed at the endpoints of the piston configuration $x=0$ and $x=1$. Transmittal boundary conditions are, instead, imposed on the piston itself at $x=a$. For a two-dimensional spherical piston $N$ of unit radius the lowest eigenvalue of the Laplacian can be found to be, from (\ref{79}), $\lambda_{0}=1/2$. This implies, according to the constraint (\ref{25}), that the allowed values of the parameter $U$ satisfy, in this case, the inequality 
\begin{equation}
U<\frac{1}{2}\frac{\sinh(1/2)}{[\sinh(1/4)]^{2}}\simeq 4.083\;.
\end{equation}   
Figure \ref{fig1} displays the Casimir force acting on the piston positioned at $x=a$ with $a\in(0,1)$. The lines of different thickness represent
the graph of the Casimir force for 
different values of the parameter $U$. Figure \ref{fig1} shows, in particular, the Casimir force on the piston when $U=\{-2,-0.8,-0.1,0.8,1.7,3.2\}$. 
The thicker the line the closer the value of $U$ for that line is to the upper limit $U_{DD}\simeq 4.083$. 

We would like to make a remark at this point. From the graphs of the Casimir force in figure \ref{fig1} one can notice that when 
the values of $U$ are negative the piston is repelled from both endpoints of the piston configuration while when $0<U<U_{DD}$ the 
piston is always attracted to the closest endpoint. This implies that by changing the sign of the parameter $U$ in the transmittal boundary condition one can change
the Casimir force from repulsive to attractive. The cutoff value is $U=0$ which represents a perfectly transparent piston. In this case the field propagates 
right through the piston and is not influenced by its presence. In this situation one has \emph{de facto} no piston configuration.

\subsection{Neumann-Neumann boundary conditions}

\begin{figure}[]
\centering
\includegraphics[scale=0.60,trim=0cm 0cm 0cm 0cm, clip=true, angle=0]{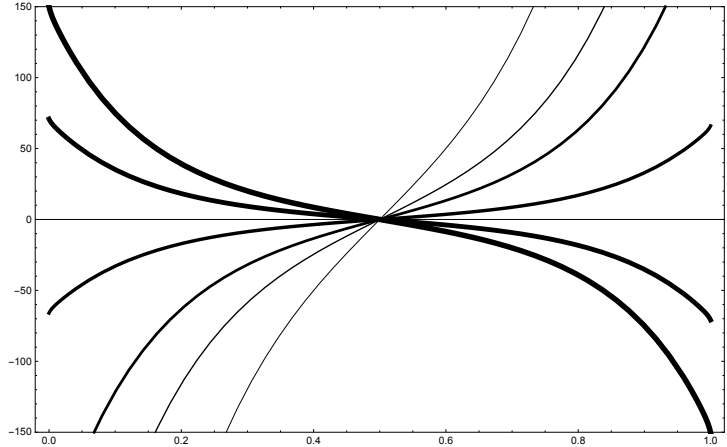}
\caption{Neumann boundary conditions at $x=0$ and $x=1$, and transmittal boundary conditions at $x=a$. The values along the $y$-axis provide the magnitude (in units for which $h=c=1$) of the Casimir force on the piston.}\label{fig2}
\end{figure}

We focus, now, on the Neumann-Neumann case. Namely, we impose Neumann boundary conditions at $x=0$ and $x=1$, and impose transmittal boundary conditions on the piston. 
In this case the parameter $U$ needs to satisfy the inequality $U<U_{NN}$ which for a spherical piston with $d=2$ and unit radius reads, according to (\ref{26}),
\begin{equation}
U_{NN}=\frac{1}{2}\tanh\left(\frac{1}{2}\right)\simeq 0.231\;.
\end{equation}  
Figure \ref{fig2} shows the Casimir force acting on the piston positioned at $x=a$ with $a\in(0,1)$ for values of $U$ in the set
$U=\{-2,-0.8,-0.4,-0.1,0.1,0.2\}$. Just like the previous case, the thicker lines represent the graph of the Casimir force on the piston for values of $U$ 
closer to the upper limit $U_{NN}\simeq 0.231$.

From the graphs in figure \ref{fig2} it is not difficult to observe that for negative values of the parameter $U$ the Casimir force tend to move the 
piston to the closest endpoint while for values of $U$ in the interval $0<U<U_{NN}$ the piston gets shifted away from the endpoints of the piston configuration.
Once again, we can conclude that also in this case a sign change in $U$ changes the attractive or repulsive nature of the force on the piston.

\subsection{Mixed boundary conditions}

\begin{figure}[]
\centering
\includegraphics[scale=0.60,trim=0cm 0cm 0cm 0cm, clip=true, angle=0]{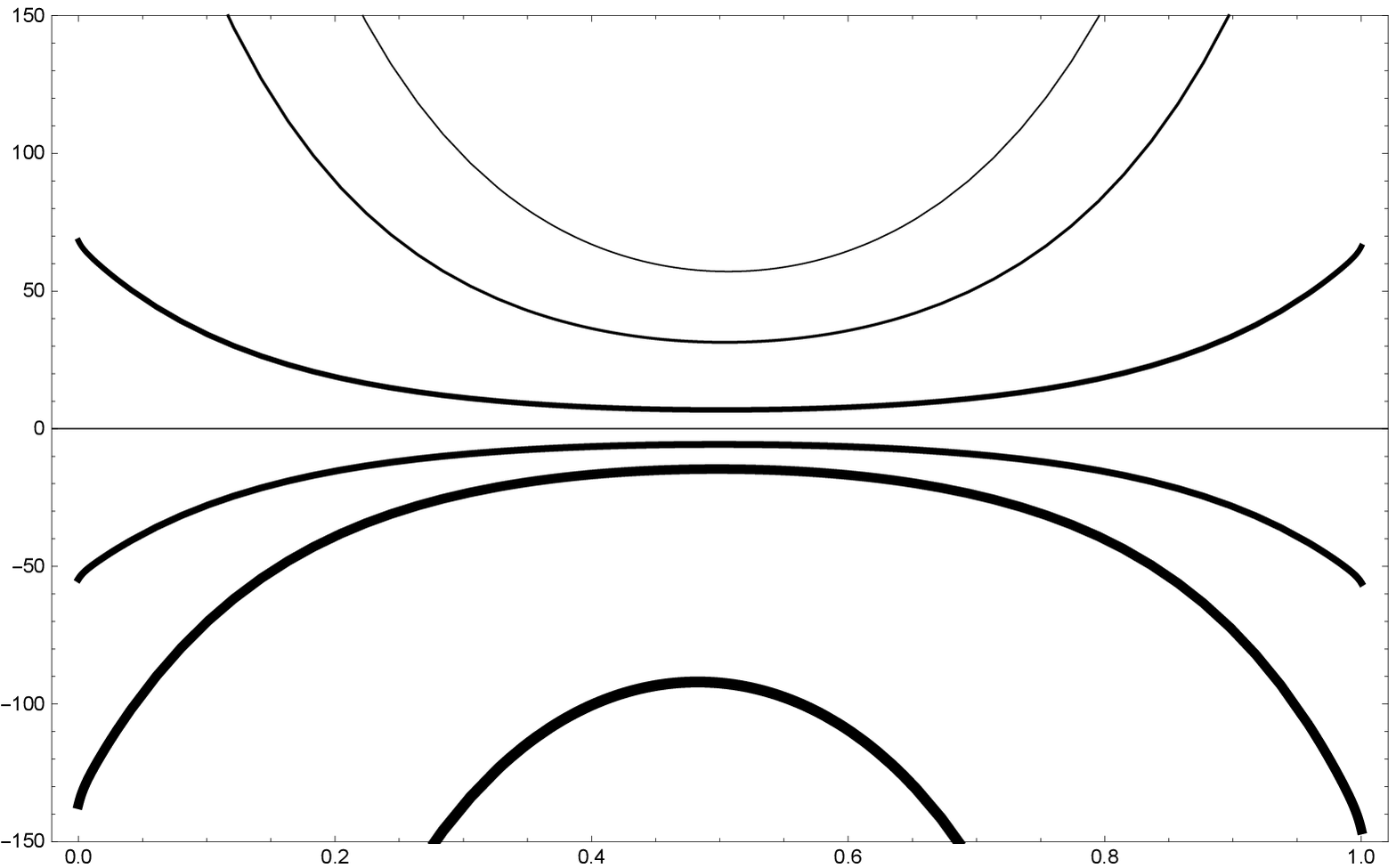}
\caption{Dirichlet boundary conditions at $x=0$ and Neuamnn boundary conditions at $x=1$. In addition, transmittal boundary conditions are imposed at $x=a$. The values along the $y$-axis provide the magnitude (in units for which $h=c=1$) of the Casimir force on the piston.}\label{fig3}
\end{figure}

\begin{figure}[]
\centering
\includegraphics[scale=0.60,trim=0cm 0cm 0cm 0cm, clip=true, angle=0]{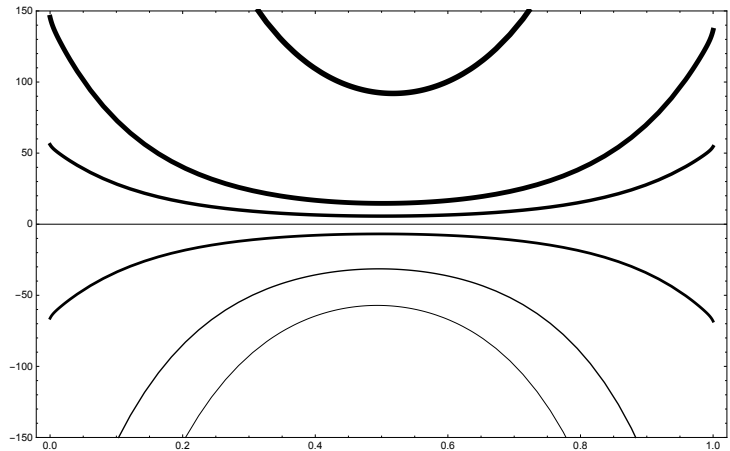}
\caption{Neumann boundary conditions at $x=0$ and Dirichlet boundary conditions at $x=1$. In addition, transmittal boundary conditions are imposed at $x=a$. The values along the $y$-axis provide the magnitude (in units for which $h=c=1$) of the Casimir force on the piston.}\label{fig4}
\end{figure}

We analyze, lastly, the case of mixed boundary conditions. This case, as already mentioned earlier, contains two types of boundary conditions. 
In one instance Dirichlet boundary conditions are imposed at $x=0$ and Neumann boundary conditions are imposed at $x=1$ and in the other the roles are reversed.
In both cases, however, transmittal boundary conditions are imposed on the piston. In both the Dirichlet-Neumann and Neumann-Dirichlet case, the parameter $U$
in the transmittal boundary conditions must satisfy the inequality $U<U_{M}$ where, according to (\ref{27}), 
\begin{equation}\label{84}
U_{M}=\frac{1}{2\tanh(1/2)}\simeq 1.082\;.
\end{equation}  
In figure \ref{fig3} we have the Casimir force on the piston in the Dirichlet-Neumann case and in figure \ref{fig4} we have the Casimir force for the Neumann-Dirichlet case.
Once again, the thicker the line the closer $U$ is to the upper limit $U_{M}$ in (\ref{84}). More precisely, the graphs, for both cases, were obtained for 
$U=\{-1,-0.5,-0.1,0.08,0.2,1\}$

As we can clearly see from the Dirichlet-Neumann graphs in figure \ref{fig3}, for negative values of $U$ the Casimir fore on the piston is always positive. 
This implies that the piston is attracted to the endpoint at $x=1$. When $0<U<U_{M}$, instead, the piston is always attracted to the $x=0$ endpoint. In the Neumann-Dirichlet 
case the behavior is exactly the opposite of the one described for the Dirichlet-Neumann case as one can easily deduce from the graphs in figure \ref{fig4}.
Once again, changing the sign on the parameter $U$ changes the sign of the Casimir force.

\section{Conclusions}

In this work we have analyzed the Casimir energy, and the corresponding force, for a massless scalar field propagating 
on a piston configuration of the type $I\times N$. The field is assumed to satisfy Dirichlet or Neumann boundary conditions at the endpoints 
of the piston configuration and transmittal boundary conditions on the pistons itself. A regularization scheme based on the spectral zeta function 
has been utilized to obtain explicit expressions for the Casimir energy and force acting on the piston. For this configuration we analyzed three 
types of boundary conditions which we denoted by Dirichlet-Dirichlet, Neumann-Neumann, and mixed. The spectral zeta function associated with the piston configuration 
has been analytically continued to a neighborhood of the point $s=-1/2$. This procedure allowed us to explicitly evaluate the 
Casimir energy and force for a general piston $N$. The general results found in this work have then been used to 
analyze the Casimir energy and force for the three types of boundary conditions in the case in which the piston is a $d$-dimensional sphere.
Numerical results have been shown for a two-dimensional spherical piston with unit radius. Obviously, by using the general formulas one could 
obtain explicit results for any given dimension $d$ and also, by suitably rescaling the spectral zeta function $\zeta_{N}(s)$, for any specified radius of the sphere.    

As we have already mentioned earlier, this work is focused on the analysis of the Casimir energy and force for a piston configuration endowed with 
transmittal boundary conditions on the piston and simple Dirichlet or Neumann boundary conditions at the endpoints. It seems natural that the next step 
in this investigation would consist in considering more general boundary conditions at the endpoints of the piston configuration. 
Such generalized boundary conditions can be written as a linear combination, through real coefficients, of the values of the field and its derivative at the given endpoint 
(see for instance \cite{fucci15}). Unfortunately, finding a suitable range of values for the real coefficients characterizing the general boundary conditions and the 
parameter $U$ that leads to a self-adjoint boundary value problem proves to be a prohibitive task within the formalism employed in this work.
It would be very interesting to understand whether the method developed in \cite{asorey13} could be more appropriate to analyze the Casimir effect 
in this more general case.      
 
A number of generalizations to this work can be envisaged. Apart from considering more general boundary conditions, as mentioned above, 
it would be very interesting to analyze piston configurations possessing different types of geometry. 
For instance, one could consider a piston configuration constructed from three concentric spheres (or three coaxial cylinders) 
with Dirichlet or Neumann boundary conditions imposed on the innermost and outermost surfaces and transmittal boundary conditions 
imposed on the one between the two. One could also consider warped piston configurations of the type $I\times_{f} N$ where, $I=[0,L]$, and $f$ is a 
suitable warping function as considered in \cite{fucci12}. In this case transmittal boundary conditions would be imposed on the piston represented by the manifold $N$ positioned 
at $a\in(0,L)$. It would be particularly intriguing to understand how the presence of both the warping function and the parameter $U$ influence the 
Casimir force on the piston and if, for a given warping function $f$, one could find non-vanishing values of $U$ for which the piston experiences no Casimir force.     
We hope to report on some of these generalizations in future works.

\end{document}